\documentclass[a4paper,12pt]{article}
\usepackage[pctex32]{graphics}

\begin{document}
\topmargin 0pt \oddsidemargin 0mm

\renewcommand{\thefootnote}{\fnsymbol{footnote}}
\begin{titlepage}
\begin{flushright}
INJE-TP-07-04
\end{flushright}

\vspace{5mm}
\begin{center}
{\Large \bf Holographic interacting  dark energy in the braneworld
cosmology } \vspace{12mm}

{\large  Kyoung Yee Kim, Hyung Won Lee  and Yun Soo
Myung\footnote{e-mail
 address: ysmyung@inje.ac.kr}}
 \\
\vspace{10mm} {\em  Institute of Mathematical Sciences and School
of Computer Aided Science \\ Inje University, Gimhae 621-749,
Korea}

\end{center}

\vspace{5mm} \centerline{{\bf{Abstract}}}
 \vspace{5mm}

We investigate  a model of  brane cosmology to find a unified description of the radiation-matter-dark energy universe.
It is  of the
interacting holographic dark energy  with a bulk-holographic matter $\chi$.
This  is a  five-dimensional cold dark matter, which plays a role of   radiation  on the brane.
Using the effective equations of state $\omega^{\rm eff}_{\rm \Lambda}$ instead of the native
equations of state $\omega_{\rm \Lambda}$, we show that
this model cannot
accommodate any transition from the dark energy with $\omega^{\rm eff}_{\rm \Lambda}\ge-1$ to the phantom
regime $\omega^{\rm eff}_{\rm \Lambda}<-1$.
Furthermore, the case of  interaction between  cold dark matter  and
five dimensional cold dark matter
 is considered for completeness. Here we find that the redshift of matter-radiation equality $z_{\rm eq}$ is
 the same order as $z^{\rm ob}_{\rm eq}=2.4\times10^{4} \Omega_{\rm m}h^2$.
 Finally, we obtain a general  decay rate $\Gamma$ which is suitable for describing  all interactions including
 the interaction between holographic dark energy and cold dark matter.

\end{titlepage}
\newpage
\renewcommand{\thefootnote}{\arabic{footnote}}
\setcounter{footnote}{0} \setcounter{page}{2}

\section{Introduction}
Recent observations from Supernova (SN Ia)~\cite{SN} and large
scale structure~\cite{SDSS} imply that our universe is
accelerating.   Also cosmic microwave background
observations~\cite{Wmap1,Wmap3} provide an evidence for the
present acceleration. A combined analysis of cosmological
observations shows that the present universe consists of 70\% dark
energy and 30\% dust matter including cold dark matter (CDM) and baryons.

Although there exist a number of dark energy models, a promising
candidate is the cosmological constant. However, one has the two
famous cosmological constant problems: the fine-tuning and
coincidence problems. In order to solve the first problem, we may introduce
a dynamical cosmological constant model inspired  by the
holographic principle.  The authors in~\cite{CKN} showed that in
quantum field theory, the UV cutoff $\Lambda$ could be related to
the IR cutoff $L_{\rm \Lambda}$ due to the limit set by
introducing  a black hole (the effects of gravity). In other
words, if $\rho_{\rm \Lambda}=\Lambda^4$ is the vacuum energy
density caused by the UV cutoff, the total energy of system with
the size $L_{\rm \Lambda}$ should not exceed the mass of the black
hole with the same size $L_{\rm \Lambda}$: $L_{\rm \Lambda}^3
\rho_{\rm \Lambda}\le 2M_p^2L_{\rm \Lambda}$. If the largest
cutoff $L_{\rm \Lambda}$ is chosen to be the one saturating this
inequality,  the holographic energy density (HDE) is  given by
$\rho_{\rm \Lambda}= 3c^2M_p^2/8\pi L_{\rm \Lambda}^2$ with a
constant $c$. The lower limit of $c$ is protected as $c\ge 1$ by the
entropy bound. Here we regard $\rho_{\rm \Lambda}$ as a dynamical
cosmological constant. Taking the IR cutoff as the size of the
present universe ($L_{\rm \Lambda}=1/H$), the resulting energy  is close to the present
dark energy~\cite{HMT}. However, this
 approach with $L_{\rm
\Lambda}=1/H$ is not fully satisfied  because it fails to recover the
equation of state (EoS) for the dark energy-dominated
universe~\cite{HSU}. Further studies in \cite{LI,FEH,Myung2,Zim3}
have shown that choosing the future event horizon as the IR cutoff
determines  an accelerating universe with the native EoS $\omega_{\rm
\Lambda} \equiv -1/3(1+ d \ln \rho_{\rm \Lambda}/ d\ln a) =-1/3-2\sqrt{\Omega_{\rm \Lambda}}/3c$.

 Also if the interaction is
turned on, the coincidence problem could be resolved~\cite{BS}.  The
interacting dark energy models provided a new direction to
understand the dark energy~\cite{Hor,Szy,Cop}. The authors in
\cite{WGA} introduced an interacting holographic dark energy model
where an interaction exists between HDE  and CDM. They derived the phantom-phase of $\omega_{\rm
\Lambda}<-1$ using  $\omega_{\rm \Lambda}$. However,
it turned out that
 the interacting holographic dark energy model could not  describe a
phantom regime  when using
the effective equation of state  $\omega^{\rm eff}_{\rm
\Lambda}$~\cite{KLM}.  More recently, it was shown  that for non-flat universe of
$k\not=0$~\cite{HL,WLA}, the interacting holographic dark energy
model could not describe a phantom regime of $\omega^{\rm
eff}_{\rm \Lambda}<-1$~\cite{Seta}.  In Ref.\cite{SV},
the authors discussed the cosmological dynamics of interacting holographic dark energy model using the phase-space
variables.  A key of this system is an interaction
between two matters. Their contents are changing due to energy
transfer from HDE to CDM until the two
components are comparable. If there exists a source/sink in the
right-hand side of the continuity equation, we must be careful to
define its EoS. In this case, the effective EoS is the only
candidate to represent the state of the mixture of two components
arisen from decaying of HDE into CDM.
This is  clearly different  from the non-interacting case which can be
described by the native EoS  $\omega_{\rm \Lambda}$ completely.

On the other hand, if the brane cosmology is introduced, one could
have interesting interaction between  bulk and brane matters. In
the low energy limit, the brane cosmology reduces to the
Friedmann-Robertson-Walker (FRW) form with   a bulk-holographic matter $\chi$.
 This is just a five dimensional cold dark matter(5DCDM) which  play a role of a four-dimensional  radiation
  when using the
effective EoS approach. Then  a unified description of radiation-matter-dark energy universe could be performed
within the  brane cosmology.
Here we obtain two kinds of interaction:
HDE-5DCDM and CDM-5DCDM.  The first
interaction may be allowed because one may allow the interaction of HDE with radiation.
However, the latter seems not to be
permitted because we assume that the CDM is not  a source of
radiation and it does not interact with the radiation. However, we
suggest that the two interactions are possible to occur within the
brane cosmology.

Concerning the brane-bulk interaction, there were contradictions: if one uses the effective EoS of
$w_{\rm de}^{\rm eff}$~~\footnote{The authors in~\cite{CGW} use a different definition $w_{\rm de}^{\rm eff}=
-1 -\frac{1}{3}\frac{d \ln (\delta H^2)}{d\ln a}$ from our definition $w_{\rm \Lambda}^{\rm eff}$.
Here $\delta H^2=H^2/H^2_0-\Omega_{\rm m}/a^3$ accounts for all terms in the
Friedmann equation not related to the brane matter $\Omega_{\rm m}$. The $w_{\rm de}^{\rm eff}=-1$ crossing
is achieved  by considering the brane-bulk interaction without specifying dark energy as holographic dark energy.},
 a transition occurs between  $w_{\rm de}^{\rm eff}>-1$
 and $w_{\rm de}^{\rm eff}<-1$~\cite{CGW}.
On the other hand, using $w_{\rm \Lambda}^{\rm eff}$, it was shown that such a transition does not occur~\cite{Setare2}.
In this work, we wish to address this issue again.
We solve three
coupled differential equations for density parameters $\Omega_{\rm
i}$ numerically by assuming
three interactions between them. Furthermore, we
introduce  three types  of the decay rate $\Gamma$ to find the  dark
energy-dominated evolution on the brane. We confirm that any
phantom-phase is not found on the brane.

\section{Brane-bulk interaction model }

Generalization of the  Randall-Sundrum scenario~\cite{RS} in cosmology
considers the AdS$_{5}$  geometry containing the bulk
cosmological constant $\Lambda$, but explores arbitrary energy
densities on the brane and in the bulk.  The
Binetruy-Deffayet-Langlois (BDL) approach is a genuine extension
of the Kaluza-Klein cosmology to account for the local
distribution on the brane~\cite{BDL}. In this case, the location of the brane
is fixed with respect to the bulk direction. This approach is
useful for describing the cosmological evolution of the brane when
a brane-bulk interaction exists.  Hence, we follow the BDL
brane cosmology.  We introduce the
gaussian-normal bulk metric for ($1+3+1$)-dimensional spacetime
\begin{equation}
\label{1eq1} ds^2_{BDL} = -c^2(t,y) dt^2
+a^2(t,y)\gamma_{ij}dx^idx^j +b^2(t,y)dy^2,
\end{equation}
where $\gamma_{ij}$ is the metric of a three-dimensional space
with a constant curvature of $6k$.  Let us  express the bulk
Einstein equation $G_{ M N}=\frac{1}{2M^3} T_{ M N}$ in terms of
the BDL metric~\footnote{Our action is given by $S_5= \int d^5x
\sqrt{-g} \big(M^3R-\Lambda + \tilde {\cal L}^{mat}_B \Big) + \int
d^4x \sqrt{-\hat g}  {\cal L}^{mat}_b $ with $ M^3=1/16 \pi
G_5=1/2\kappa^2_5$ and ${\cal L}^{mat}_b=- (\sigma + \rho)$~\cite{MK}. }. We
introduce a $(1+3)$-dimensional brane located at $y=0$. For
simplicity,   we choose
 the total stress-energy tensor $ T^M~_N={\rm
diag}(-\Lambda,-\Lambda,-\Lambda,-\Lambda,-\Lambda)+ \tilde T^M~_N
+\tau^{\mu}~_\nu$. Here $\Lambda$ is  the bulk cosmological
constant and the bulk stress-energy tensor $\tilde T^M~_N$ from
$\tilde {\cal L}^{mat}_B$ is not needed to have a specific form
initially. If $\tilde T^t~_y=0$, it is obvious that there is no
brane-bulk interaction. The brane stress-energy
tensor from ${\cal L}^{mat}_b$ including the brane tension
$\sigma$ and the energy density $\rho$ is assumed to take  the
form
\begin{equation}
\label{1eq2}  \tau^{\mu}~_\nu = \frac{\delta(y)}{b}{\rm
diag}(-\rho-\sigma,p-\sigma,p-\sigma,p-\sigma,0).
\end{equation}
  We are interested in
solving the Einstein equations at the location of the brane. Initially we
indicate by the subscript ``0" for the value of various quantities
on the brane. Also it is convenient to choose the
gaussian-normal gauge with $b_0=1$ and the temporal gauge with
$c_0=1$ on the brane. We obtain from $G_{0y}=\frac{1}{2M^3}
T_{0y}$,
\begin{equation}
\label{1eq3} \dot \rho +3 \frac{\dot a_0}{a_0}\rho(1+\omega)= -2
\tilde T^0~_y.
\end{equation}
Here we assumed an equation of state $p=\omega \rho$ on the brane.

 On the other
hand, the average part of $yy$-component equation is given by
\begin{equation}
\label{1eq4}
 \frac{\ddot a_0}{a_0} + \Big( \frac{\dot a_0}{a_0} \Big )^2 +
 \frac{k}{a_0^2} =\frac{1}{6M^3} \Big( \Lambda + \frac{\sigma^2}{12 M^3}
 \Big) -\frac{1}{144M^6}\Big(\sigma(3p -\rho)+\rho(3p+\rho)\Big)-
 \frac{1}{6M^3} \tilde T^y~_y.
  \end{equation}
 Then, we rewrite
Eq.(\ref{1eq4}) in the following equivalent form  by introducing
the two bulk-holographic energy densities $\tilde{\chi}$ and $\phi$:
\begin{eqnarray}
\label{1eq5} && H^2_0 = \frac{1}{144M^6} \Big( \rho^2 + 2\sigma
\rho \Big)  + \tilde{\chi} +\phi+ \frac{1}{12M^3}
\Big( \Lambda + \frac{\sigma^2}{12 M^3}
 \Big) -\frac{k}{a^2_0},  \\
 \label{1eq6}
 &&\dot {\tilde{\chi}}+4 H_0 \tilde{\chi} = \frac{1}{36M^6} (\rho+\sigma) \tilde
 T^0~_y,\\
  \label{1eq7}
 && \dot \phi +4 H_0 \phi =- \frac{1}{3M^3} H_0 \tilde T^y~_y ,
\end{eqnarray}
with $H_0=\dot a_0/a_0$. In the case of $p=\rho=0$ and
$\phi=\tilde{\chi}=0$, one finds  the Randall-Sundrum vacuum state~\cite{RS}.
We choose the cosmological constant $\Lambda = -
\sigma^2 / 12M^3 = -12M^3 / \ell^2$ with  the brane tension
$\sigma=12 M^3 /\ell$ to have a critical brane. Hence the
cosmological evolution will be determined by four initial
parameters $(\rho_i,a_{0i},\tilde{\chi}_i,\phi_i)$ instead of
two $(\rho_i,a_{0i})$ in the FRW
universe. This is so because the generalized Friedmann equation
(\ref{1eq5}) is not a first integral of the Einstein equation. It
is mainly due to the energy exchange $\tilde T^t~_y $ between the
brane and  bulk. In the case of $\phi=0$ and $\tilde T^t~_y=A \rho>0$
with $\tilde T^y~_y =0$, one finds a mirage-radiation term
$\tilde{\chi} \sim (1-e^{-At/2})/a_{0}^4$ for an energy outflow
from the brane~\cite{KKTTZ}. It is a cosmological model that the
real matter on the brane decays into the extra dimension.  Also for
$\phi=0, \tilde T^t~_y \sim -\frac{1}{a_0^q}$ and
$\tilde T^y~_y=0$, it is shown that the energy influx from the bulk
generates a cosmological acceleration on the brane with the
acceleration parameter $Q \equiv\frac{1}{H_0^2} \frac{\ddot
a_0}{a_0}=1-\frac{q}{3}$, where $0 \le q \le 4$ \cite{Tet}.
However, in general, it will be a formidable task to solve
Eqs.(\ref{1eq5})-(\ref{1eq7}) with Eq.(\ref{1eq3})  because it
gives rise to a complicated dynamics between the brane and the
bulk. In  Ref.\cite{CGW}, they used  $ \tilde T^t~_y \propto
Ha^n$ to derive the super-acceleration using $\omega_{\rm de}^{\rm eff}$.

For our purpose,  let us imagine a brane universe made of CDM
$\rho_{\rm m}$ with $\omega_{\rm m}=0$, but obeying the
holographic principle. In addition, we propose that the
holographic energy density $\rho_{\rm \Lambda}$ exists with
its native EoS $\omega_{\rm \Lambda}\ge-1$ on the brane. If one assumes a form of the
interaction $T$ with
$\phi=\tilde{T}^y~_y=0$, their continuity equations take the
simple forms\footnote{Hereafter, we focus on the brane. Hence we
use the notation without the subscript ``0" and
$T=2\tilde{T}^0~_{y}$, and $\chi=(72M^6/\sigma)\tilde{\chi}$, and
$\sigma/77M^6=1/6M^3 \ell=8 \pi/3M_p^2$. Also we concentrate on the low-energy region of $\rho \ll \sigma$
and thus $\rho^2$-term in Eq.(\ref{1eq5}) is negligible. }
\begin{eqnarray}
\label{2eq1}&& \dot{\rho}+3H(1+\omega)\rho =-T, ~~\rho=\rho_{\rm \Lambda}+\rho_{\rm m}\\
\label{2eq2}&& \dot{\chi}+4H\chi=T
\end{eqnarray}
and the generalized Friedmann equation (\ref{1eq5}) on the
critical brane  leads to
\begin{equation}
\label{2eq3} H^2=\frac{8\pi}{3M^2_p}\Big[
 \rho+\chi\Big]-\frac{k}{a^2}.
\end{equation}
Now we consider the case of decaying from  HDE to 5DCDM  with $T=\Gamma \rho_{\rm \Lambda}$,
while the CDM is conserved by choosing
\begin{eqnarray}
\label{2eq4}&& \dot{\rho}_{\rm \Lambda}+3H(1+\omega_{\rm \Lambda})\rho_{\rm \Lambda} =-T,\\
\label{2eq5}&& \dot{\rho_{\rm m}}+3H\rho_{\rm m}=0.
\end{eqnarray}
This decaying process impacts
their equations of state  and particularly, it induces the effective EoS for the 5DCDM.
Interestingly, an  accelerating
phase could arise from a large effective non-equilibrium pressure
 $\Pi_{\rm \chi}$  defined as $\Pi_{\rm \chi}\equiv -\Gamma\rho_{\rm \Lambda}/3H(=\Pi_{\rm
 \Lambda})$. Then  the two  equations (\ref{2eq4})
and (\ref{2eq2}) are translated into those of the two
dissipatively imperfect fluids
\begin{eqnarray}
\label{2eq6}&& \dot{\rho}_{\rm \Lambda}+ 3H\Big[1+\omega_{\rm
\Lambda}+\frac{\Gamma}{3H} \Big]\rho_{\rm \Lambda}=\dot{\rho}_{\rm
\Lambda}+ 3H\Big[(1+\omega_{\rm \Lambda})\rho_{\rm
\Lambda}+\Pi_{\rm
 \Lambda}\Big]=0, \\
\label{2eq7}&& \dot{\chi}+3H\Big[1+\frac{1}{3}-\frac{\rho_{\rm \Lambda}}{
\chi}\frac{\Gamma}{3H}\Big]\chi=\dot{\chi}+3H\Big[(1+\frac{1}{3})\chi-\Pi_{\rm \chi}\Big]=0.
\end{eqnarray}
The positivity of $\Pi_{\rm \Lambda}>0$ shows a decaying  of
HDE via the cosmic frictional force, while
$\Pi_{\rm \chi}<0$ induces a production of the mixture via the cosmic
anti-frictional force simultaneously~\cite{Zim1,myung}. This is a
sort  of the vacuum decay process to generate a particle
production within the two-fluid model~\cite{Zim2}. As a result, a
mixture of two components will be  created. When turning on the interaction term, from Eqs.(\ref{2eq6})
and (\ref{2eq7}), we read off their
effective equations of state as
\begin{equation}
\label{2eq8} \omega^{\rm eff}_{\rm \Lambda}=\omega_{\rm
\Lambda}+\frac{\Gamma}{3H},~~ \omega^{\rm eff}_{\rm
\chi}=\frac{1}{3}-\frac{\rho_{\rm \Lambda}}{\chi}\frac{\Gamma}{3H}. \end{equation}
Hence it is clear that the 5DCDM $\chi$ plays a role of  radiation on the brane, if there is no interaction.
Introducing the density parameters defined by $\Omega_{\rm i}=\rho_{\rm i}/\rho_{\rm c}$ as
\begin{equation}
\label{2eq9} ~\Omega_{\rm m}=\frac{8\pi \rho_{\rm
m}}{3M_p^2H^2},~\Omega_{\rm \Lambda}=\frac{8 \pi \rho_{\rm
\Lambda}}{3M^2_pH^2},~ \Omega_{\rm k}=\frac{k}{a^2H^2},~~
\Omega_{\rm \chi}=\frac{8\pi \chi}{3M_p^2H^2},
\end{equation}
we can  rewrite the Friedmann equation (\ref{2eq3}) as a simplified form

\begin{equation} \label{2eq10} \Omega_{\rm m}+\Omega_{\rm
\Lambda}+\Omega_{\rm \chi}=1+\Omega_{\rm k}.\end{equation}
Hereafter we use this relation instead of Eq.(\ref{2eq3}).

For the non-flat universe of $k \not=0$, we  introduce the future event horizon $L_{\rm
\Lambda}=R_{\rm FH}=a\xi_{\rm FH}(t)=a \xi^{k}_{\rm FH}(t)$
with
\begin{equation}
\label{2eq11} \xi_{\rm FH}(t)=\int_t^{\infty} \frac{dt}{a}.
\end{equation}
Here the comoving horizon size is given by
\begin{equation} \label{2eq12}
\xi^k_{\rm
FH}(t)=\int_{0}^{r(t)}\frac{dr}{\sqrt{1-kr^2}}=\frac{1}{\sqrt{|k|}}{\rm
sinn}^{-1}\Bigg[\sqrt{|k|}r(t)\Bigg],
\end{equation}
where leads to $\xi^{k=1}_{\rm FH}(t)={\rm sin}^{-1}r(t)$,
$\xi^{k=0}_{\rm FH}(t)=r(t)$, and  $\xi^{k=-1}_{\rm FH}(t)={\rm
sinh}^{-1}r(t)$.  For our purpose, we use a comoving radial
coordinate $r(t)$,
\begin{equation} \label{2eq13}
r(t)=\frac{1}{\sqrt{|k|}} {\rm sinn}\Bigg[\sqrt{|k|}\xi^{k}_{\rm
FH}(t)\Bigg].
\end{equation}
 $L_{\rm \Lambda}=ar(t)$ is
a useful length scale  for the non-flat universe~\cite{HL}. Its derivative with respect to time $t$ leads to
\begin{equation} \label{2eq14}
\dot{L}_{\rm \Lambda}=H L_{\rm \Lambda}+a
\dot{r}=\frac{c}{\sqrt{\Omega_{\rm \Lambda}}}-{\rm cosn}y,
\end{equation}
where ${\rm cosn}y={\rm cos}y,~y,~{\rm cosh}y$ for $k=1,0,-1$ with $y=\sqrt{k} R_{\rm FH} /a$. Hereafter we consider three classes
of interactions: HDE-CDM, HDE-5DCDM, and CDM-5DCDM.
Using the definition of $\rho_{\rm
\Lambda}=\frac{3c^2M_p^2}{8\pi L_{\rm \Lambda}^2}$ and
(\ref{2eq8}), one finds the  equation of state for HDE

\begin{equation} \label{3eq1}
\dot{\rho}_{\rm \Lambda} +3H \Big[1-
\frac{1}{3}-\frac{2\sqrt{\Omega_{\rm \Lambda}}}{3c}{\rm
cosn}y\Big]\rho_{\rm \Lambda}=0.
\end{equation}
Here we can read off the effective EoS for HDE as
\begin{equation}
\label{3eq2} \omega^{\rm eff}_{\rm \Lambda}(x)=
-\frac{1}{3}-\frac{2\sqrt{\Omega_{\rm \Lambda}(x)}}{3c}{\rm cosn}y
\end{equation}
with $x=\ln a$.
At the first sight, the above effective EoS seems not to be relevant to the interaction,
but it depends on the decay rate $\Gamma$ through $\Omega_{\rm \Lambda}$.

\section{Unified picture for interactions}
For the  interaction between   HDE  and CDM on the brane, we assume to have
\begin{eqnarray}
&&\dot \rho_{\rm m} + 3H \rho_{\rm m} = \hat{T},\\
&&\dot \rho_{\rm \Lambda} + 3H (1+\omega_{\rm \Lambda} )\rho_{\rm \Lambda} = -\hat{T}, \\
&&\dot \chi + 4H \chi = 0,
\end{eqnarray}
where $\hat{T}$ is  chosen as $\hat{T} = \Gamma \rho_{\rm \Lambda} $
for decaying from HDE to CDM, while $\hat{T} = -\Gamma \rho_{\rm m}$ for decaying from CDM to HDE.
This case is not realized by the brane cosmology because the interaction $\hat{T}$ is effective on the brane.
Hence there is no brane-bulk interaction ($T=0$).
However, we include this type of interaction for completeness.

In the case of interaction between HDE and 5DCDM, their continuity equations are given by
\begin{eqnarray}
&&\dot \rho_{\rm m} + 3H \rho_{\rm m} = 0,\\
&&\dot \rho_{\rm \Lambda} + 3H (1+\omega_{\rm \Lambda} )\rho_{\rm \Lambda} = -T, \\
&&\dot \chi + 4H \chi = T,
\end{eqnarray}
Here $T$ is chosen as $T = \Gamma \rho_{\rm \Lambda} $
for decaying from HDE to 5DCDM, whereas $T = -\Gamma \chi$ for decaying from 5DCDM to HDE.

Finally, the case of interaction between  CDM and 5DCDM takes the form
\begin{eqnarray}
&&\dot \rho_{\rm m} + 3H \rho_{\rm m} = -T,\\
&&\dot \rho_{\rm \Lambda} + 3H (1+\omega_{\rm \Lambda} )\rho_{\rm \Lambda} = 0, \\
&&\dot \chi + 4H \chi = T,
\end{eqnarray}
where  $T$ is  chosen as $T = \Gamma \rho_{\rm m} $
for decaying from CDM to 5DCDM, while  $T = -\Gamma \chi$ for decaying from 5DCDM to CDM.

By  choosing appropriate effective equations of state,
the above equations for all three cases can be unified as follows:
\begin{eqnarray}
&&\dot \rho_{\rm m} + 3H (1+\omega^{\rm eff}_{\rm m} ) \rho_{\rm m} = 0,\\
&&\dot \rho_{\rm \Lambda} + 3H (1+\omega^{\rm eff}_{\rm \Lambda} )\rho_{\rm \Lambda} = 0, \\
&&\dot \chi + 3H (1+\omega^{\rm eff}_{\rm \chi} )\chi = 0,
\end{eqnarray}
All effective EoS are summarized on the Table \ref{Tab:parameters}.
\begin{table}
\caption{\label{Tab:parameters}Summary of effective equations of state and related information.
Here I, ${\rm \Lambda}$, ${\rm m}$,
and ${\rm \chi}$ represent interaction, HDE, CDM, and 5DCDM, respectively. The redshift factor $z_{\rm eq}$ is determined from the relation
$x=-\ln(1+z)$ when $\Omega_{\rm m}=\Omega_{\rm \chi}$. NA denotes ``not available".
Finally, yes (no) represent the status of evolution.  }

\begin{tabular}{llllllll} \hline
 IT & $\omega^{\rm eff}_{\rm m}$  & $\omega^{\rm eff}_{\rm \chi}$ & $T/\Gamma$  & $\Gamma/3Hb^2$ & $z_{\rm eq}$ & status & figure \\ \hline

 no &0& $\frac{1}{3}$& 0 & 0 & 27.1 & yes & Fig. 1  \\ \hline

 ${\rm \Lambda} \rightarrow {\rm m}$ &$-\frac{\Gamma}{3H} \frac{\Omega_{\rm \Lambda}}{\Omega_{\rm m}}$
                                     & $\frac{1}{3}$& $\rho_{\rm \Lambda}$ & $( 1+ \frac{\Omega_{\rm m}}{\Omega_{\rm \Lambda}})$ &2.7& yes & Fig. 2a  \\

  &
  &
  &
  & $( 1+ \frac{\Omega_{\rm m}}{\Omega_{\rm \Lambda}})\Omega_{\rm \Lambda}$ & 10.2 & yes & Fig. 2c \\

  &&&& $( 1+ \frac{\Omega_{\rm m}}{\Omega_{\rm \Lambda}})\Omega_{\rm \Lambda}\Omega_{\rm m}$&18.9&yes&Fig. 2e  \\ \hline

 ${\rm m} \rightarrow {\rm \Lambda}$
         & $\frac{\Gamma}{3H}$
         & $\frac{1}{3}$
         & $- \rho_{\rm m}$ & $( 1+ \frac{\Omega_{\rm \Lambda}}{\Omega_{\rm m}})$ &NA&no&Fig. 2b  \\
&&&& $( 1+ \frac{\Omega_{\rm \Lambda}}{\Omega_{\rm m}})\Omega_{\rm m}$ &762.3&yes&Fig. 2d \\
&&&& $( 1+ \frac{\Omega_{\rm \Lambda}}{\Omega_{\rm m}})\Omega_{\rm m}\Omega_{\rm \Lambda}$ &36.5&yes&Fig. 2f \\ \hline

${\rm \Lambda} \rightarrow {\rm \chi}$
        &0
        &$ \frac{1}{3}- \frac{\Gamma}{3H} \frac{\Omega_{\rm \Lambda}}{\Omega_{\rm \chi}}$
        & $\rho_{\rm \Lambda}$ & $( 1+ \frac{\Omega_{\rm \chi}}{\Omega_{\rm \Lambda}})$ &NA&no&Fig. 3a\\
&&&& $( 1+ \frac{\Omega_{\rm \chi}}{\Omega_{\rm \Lambda}})\Omega_{\rm \Lambda}$ &NA&no&Fig. 3c \\
&&&& $( 1+ \frac{\Omega_{\rm \chi}}{\Omega_{\rm \Lambda}})\Omega_{\rm \Lambda}\Omega_{\rm \chi}$ &31.7&yes&Fig. 3e \\ \hline

${\rm \chi} \rightarrow {\rm \Lambda}$
        &0
        &$\frac{1}{3}+\frac{\Gamma}{3H} $
        & $-\chi$ & $( 1+ \frac{\Omega_{\rm \Lambda}}{\Omega_{\rm \chi}})$ &NA&no&Fig. 3b \\
&&&& $( 1+ \frac{\Omega_{\rm \Lambda}}{\Omega_{\rm \chi}})\Omega_{\rm \chi}$ &14.1&yes&Fig. 3d \\
&&&& $( 1+ \frac{\Omega_{\rm \Lambda}}{\Omega_{\rm \chi}})\Omega_{\rm \chi}\Omega_{\rm \Lambda}$ &22.9&yes&Fig. 3f \\ \hline

${\rm m} \rightarrow {\rm \chi}$
           &$\frac{\Gamma}{3H}$
           &$ \frac{1}{3}- \frac{\Gamma}{3H} \frac{\Omega_{\rm m}}{\Omega_{\rm \chi}}$
           & $\rho_{\rm m}$ & $( 1+ \frac{\Omega_{\rm \chi}}{\Omega_{\rm m}})$ &NA&no&Fig. 4a \\
&&&& $( 1+ \frac{\Omega_{\rm \chi}}{\Omega_{\rm m}})\Omega_{\rm m}$ &NA&no&Fig. 4c \\
&&&& $( 1+ \frac{\Omega_{\rm \chi}}{\Omega_{\rm m}})\Omega_{\rm m}\Omega_{\rm \chi}$ &1109.5&yes&Fig. 4e\\ \hline

${\rm \chi} \rightarrow {\rm m}$
           &$-\frac{\Gamma}{3H}\frac{\Omega_{\rm \chi}}{\Omega_{\rm m}}$
           &$\frac{1}{3}+\frac{\Gamma}{3H} $
           & $-\chi$ & $( 1+ \frac{\Omega_{\rm m}}{\Omega_{\rm \chi}})$ &NA&no&Fig. 4b\\
&&&& $( 1+ \frac{\Omega_{\rm m}}{\Omega_{\rm \chi}})\Omega_{\rm \chi}$ &NA&no&Fig. 4d \\
&&&& $( 1+ \frac{\Omega_{\rm m}}{\Omega_{\rm \chi}})\Omega_{\rm \chi}\Omega_{\rm m}$&10.0&yes&Fig. 4f \\ \hline
\end{tabular}
\end{table}
\begin{figure}[t!]
   \centering
\scalebox{.7}
   {\includegraphics{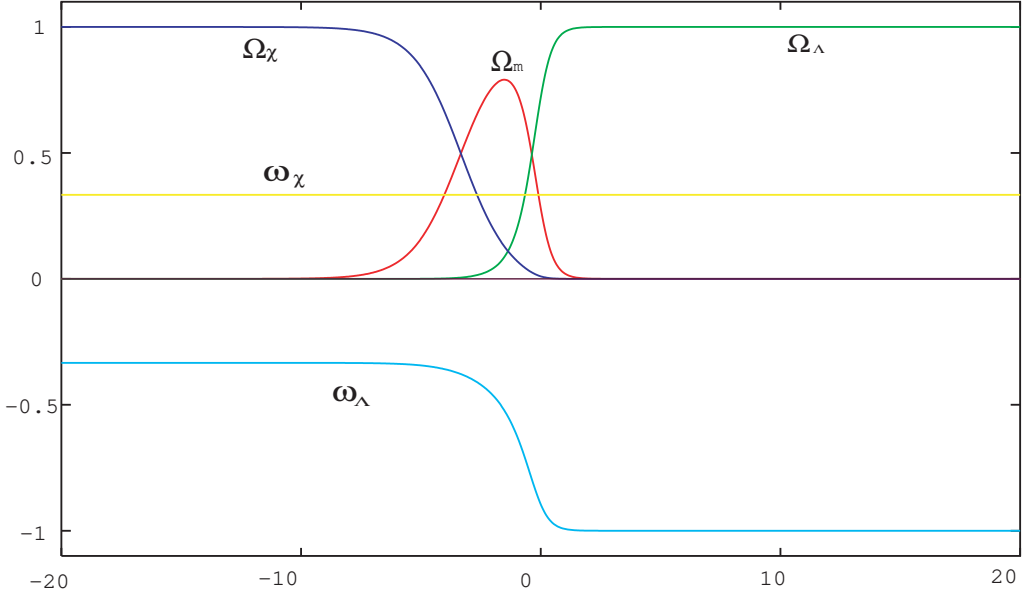}}
\caption{(color online) Graph for the noninteracting case. For $b^2=0$ and $c=1$, $k=1$ evolution
of $\Omega_{\rm \Lambda}$ (green), $\Omega_{\rm m}$ (red), and $\Omega_{\rm \chi}$ (blue) and
the  equations of state, $\omega_{\rm \Lambda}$ (cyan) and
$\omega^{\rm eff}_{\rm \chi}=1/3$ (yellow) with $\omega_{\rm m}=0$. Here $x=\ln
a$ moves backward direction ($-$) or forward direction ($+$), starting at the present time
$x=0(a_0=1)$. } \label{fig1}
\end{figure}
However, $\omega^{\rm eff}_{\rm \Lambda}$
is the same for all cases as is given by Eq.(\ref{3eq2}).
Here we choose three types for the decay rate ${\rm \Gamma}$ with $b^2=0.2$:
\begin{eqnarray}
&&{\rm (1)-type}: {\rm \Gamma}= 3Hb^2( 1+ \frac{\Omega_{\rm i}}{\Omega_{\rm j}}),\\
&&{\rm (2)-type}: {\rm \Gamma}=3Hb^2( 1+ \frac{\Omega_{\rm i}}{\Omega_{\rm j}})\Omega_{\rm j},\\
&&{\rm (3)-type}: {\rm \Gamma}=3Hb^2( 1+ \frac{\Omega_{\rm i}}{\Omega_{\rm j}})\Omega_{\rm i}\Omega_{\rm j}.
\end{eqnarray}
(1)-type is known as a conventional form for the interaction between  HDE and CDM.
However, choosing this form leads to an unwanted evolution and thus we have to introduce another
interaction (2)-type for the evolution of the dark energy-dominated universe.
Finally, (3)-type is chosen because (2)-type is not suitable for describing the interaction between CDM and 5DCDM.
Another types are found in Ref.\cite{BS}.

In order to obtain differential equations for density parameters, $\Omega_{\rm m}, \Omega_{\rm \Lambda}$ and
$\Omega_{\rm \chi}$ which govern evolution of the universe, we introduce
\begin{equation}
R_{\rm i} = \frac{\rho_i}{\rho_c} = \Omega_{\rm i},  {\rm i} = {\rm m}, {\rm \Lambda}, {\rm \chi}.
\end{equation}
Differentiating $R_{\rm i}$ with respect to cosmic time $t$ and then using appropriate definitions,
 we obtain three  equations
\begin{eqnarray}
\Omega_{\rm m}^{\prime} &=& \Omega_{\rm m} \left [
        2 + (1 +3 \omega^{\rm eff}_{\rm m}) \Omega_{\rm m}
          + (1 +3 \omega^{\rm eff}_{\rm \Lambda}) \Omega_{\rm \Lambda}
          + (1 +3 \omega^{\rm eff}_{\rm \chi}) \Omega_{\rm \chi}
              \right ] - 3 \Omega_{\rm m}(1+\omega^{\rm eff}_{\rm m}), \\
\Omega_{\rm \Lambda}^{\prime} &=& \Omega_{\rm \Lambda} \left [
        2 + (1 +3 \omega^{\rm eff}_{\rm m}) \Omega_{\rm m}
          + (1 +3 \omega^{\rm eff}_{\rm \Lambda}) \Omega_{\rm \Lambda}
          + (1 +3 \omega^{\rm eff}_{\rm \chi}) \Omega_{\rm \chi}
              \right ] - 3 \Omega_{\rm \Lambda}(1+\omega^{\rm eff}_{\rm \Lambda}), \\
\Omega_{\rm \chi}^{\prime} &=& \Omega_{\rm \chi} \left [
        2 + (1 +3 \omega^{\rm eff}_{\rm m}) \Omega_{\rm m}
          + (1 +3 \omega^{\rm eff}_{\rm \Lambda}) \Omega_{\rm \Lambda}
          + (1 +3 \omega^{\rm eff}_{\rm \chi}) \Omega_{\rm \chi}
              \right ] - 3 \Omega_{\rm \chi}(1+\omega^{\rm eff}_{\rm \chi}),
\end{eqnarray}
where $^{\prime}$ is the differentiation with respect to  $x = \ln a$. These equations
come from the first and second Friedmann equations combined with their continuity equations.
In order to obtain solution, we have to solve the above coupled
equations numerically by considering the initial condition at
present time\footnote{Here we use the data from the combination of
WMAP3 plus the HST key project constraint on $H_0$~\cite{Wmap3}.}:
$\Omega_{\rm \Lambda}^{\prime}|_{x=0}>0,~\Omega^{0}_{\rm
\Lambda}=0.72, \Omega^{0}_{\rm k=1}=0.01, \Omega^{0}_{\rm
m}=0.28, \Omega^{0}_{\rm \chi}=0.01$.

The noninteracting case with $b^2=0$ is depicted at Fig. 1, which shows the standard evolution for the
HDE.
Here  the effective EoS reduces to the native EoS because of the absence of interactions
except $\omega^{\rm eff}_{\rm \chi}=1/3$ for 5DCDM $\chi$.
Each matter satisfies its continuity equation. We find a sequence of dominance in the evolution of the universe:
radiation$\to$ CDM $\to$dark energy. The redshift factor $z_{\rm eq}=27.1$ is determined from the relation
of $x=-\ln(1+z)$ when $\Omega_{\rm m}=\Omega_{\rm \chi}$.

Figs. 2a-f show the evolution for the interaction between HDE and CDM on the brane.
The left column of HDE$\to$CDM was already known but the right column shows new results.
These all indicate  evolutions for dark energy-dominated universe except the case of CDM$\to$HDE
with the  decay rate $\Gamma=3Hb^2(1+\Omega_{\rm \Lambda}/\Omega_{\rm m})$~\cite{WGA,KLM}.
This case provides a negative
density parameter $\Omega_{\rm m}<0$
for the future evolution and thus induces the unwanted case of  $\Omega_{\rm \Lambda}>1$.

\begin{figure}[t!]
   \centering
 \scalebox{.9}
   {\includegraphics{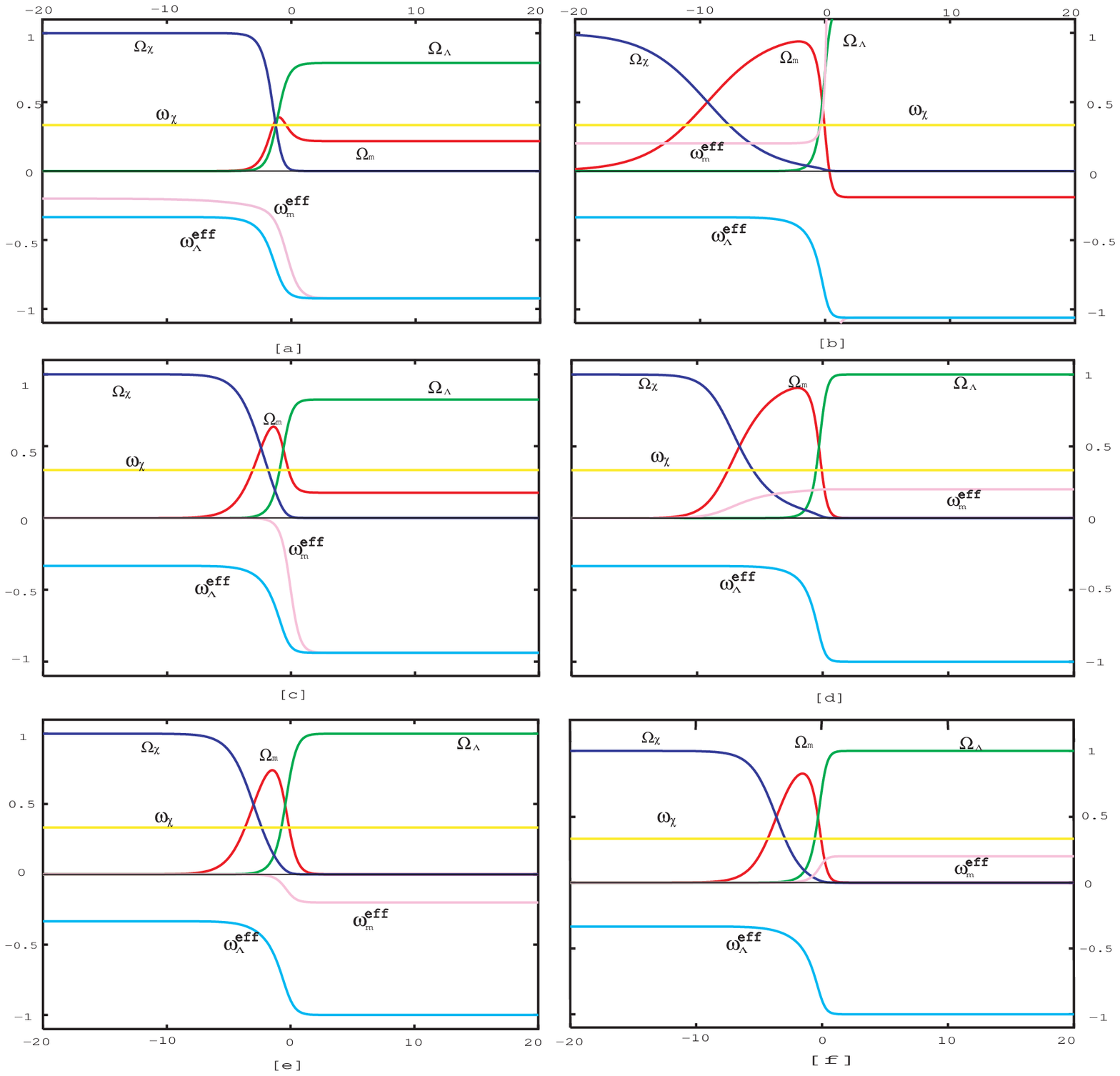}}
\caption{(color online) Six graphs for the interaction between HDE and CDM.
For $b^2=0.2$ and $c=1$, $k=1$ evolution
of $\Omega_{\rm \Lambda}$ (green), $\Omega_{\rm m}$ (red), and $\Omega_{\rm \chi}$ (blue) and
the effective equations of state, $\omega^{\rm eff}_{\rm \Lambda}$ (cyan) and
$\omega^{\rm eff}_{\rm \chi}=1/3$ (yellow) with $\omega^{\rm eff}_{\rm m}$(pink). The left
column is for HDE$\to$CDM and the right one is for CDM$\to$HDE. Fig. 2a and 2b are for the
decay rate of (1)-type, Fig. 2c and 2d  for the
decay rate of (2)-type, and Fig. 2e and 2f for the
decay rate of (3)-type.} \label{fig2}
\end{figure}
\begin{figure}[t!]
   \centering
\scalebox{.9}
   {\includegraphics{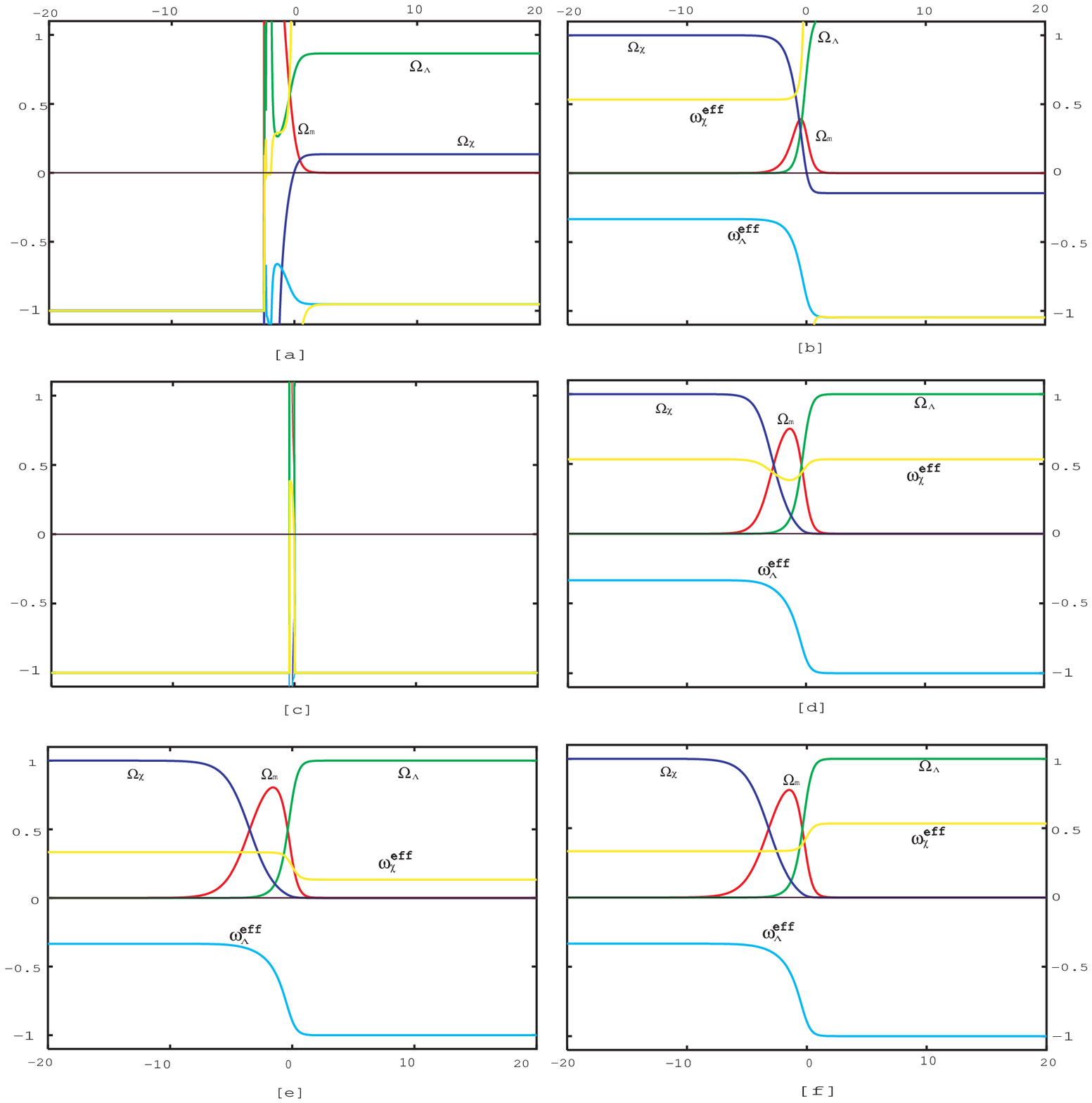}}
\caption{(color online) Six graphs for the interaction between HDE and 5DCDM.
For $b^2=0.2$ and $c=1$, $k=1$ evolution
of $\Omega_{\rm \Lambda}$ (green), $\Omega_{\rm m}$ (red), and $\Omega_{\rm \chi}$ (blue) and
the effective equations of state, $\omega^{\rm eff}_{\rm \Lambda}$ (cyan) and
$\omega^{\rm eff}_{\rm \chi}$ (yellow) with $\omega^{\rm eff}_{\rm m}=0$.  The left
column is for HDE$\to$5DCDM and the right one is for 5DCDM$\to$HDE. Fig. 3a and 3b are for the
decay rate of (1)-type, Fig. 3c and 3d  for the
decay rate of (2)-type, and Fig. 3e and 3f for the
decay rate of (3)-type. } \label{fig3}
\end{figure}
\begin{figure}[t!]
   \centering
\scalebox{.9}
   {\includegraphics{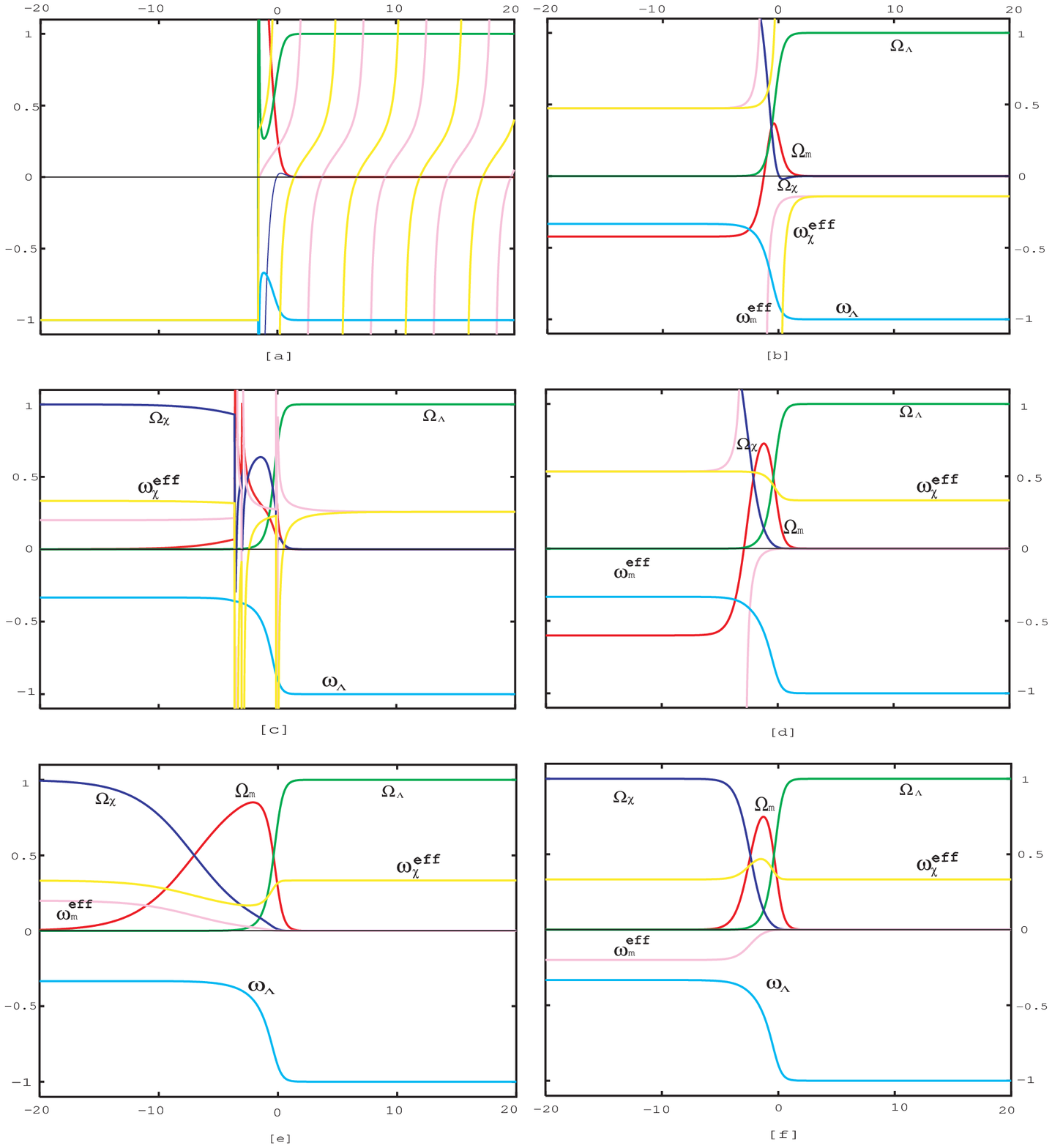}}
\caption{(color online)  Six graphs for the interaction between CDM and 5DCDM.
For $b^2=0.2$ and $c=1$, $k=1$ evolution
of $\Omega_{\rm \Lambda}$ (green), $\Omega_{\rm m}$ (red), and $\Omega_{\rm \chi}$ (blue) and
the effective equations of state, $\omega^{\rm eff}_{\rm \Lambda}$ (cyan) and
$\omega^{\rm eff}_{\rm \chi}$ (yellow) with $\omega^{\rm eff}_{\rm m}$ (pink).  The left
column is for CDM$\to$5DCDM and the right one is for 5DCDM$\to$CDM. Fig. 4a and 4b are for the
decay rate of (1)-type, Fig. 4c and 4d for the
decay rate of (2)-type, and Fig. 4e and 4f for the
decay rate of (3)-type. } \label{fig4}
\end{figure}

Figs. 3a-f indicate the evolution for the interaction between HDE and 5DCDM.
This corresponds to the case of interaction between HDE and radiation on the brane.
The left column is for  HDE$\to$ 5DCDM. An evolution for dark energy-dominated universe is possible for
only the decay rate of (3)-type: $\Gamma=3Hb^2(1+\Omega_{\rm \Lambda}/\Omega_{\rm m})\Omega_{\rm \Lambda}\Omega_{\rm m}$.
The right column is for 5DCDM$\to$HDE. Here  evolutions come out when choosing (2)and (3)-type.
All forward evolutions are possible, whereas  backward evolutions are not possible for (1)-type and  HDE$\to$ 5DCDM
with (2)-type.

Figs. 4a-f show the evolution for the interaction between CDM  and 5DCDM.
This corresponds to the case of interaction between CDM  and radiation on the brane.
The left column is for  CDM$\to$ 5DCDM and the right column is for 5DCDM$\to$CDM.
An evolution for dark energy-dominated universe is possible for
only the decay rate of (3)-type.
All backward evolutions seem not to be possible for (1) and (2)-types.
Especially, we find  the unwanted backward evolution  of $\Omega_{\rm m}<0,~\Omega_{\rm \chi}>1$ for
the 5DCDM$\to$CDM with (1) and (2)-types.
In this sense,  (3)-type is considered as the general form of decay rate $\Gamma$.

\section{Discussions}
We investigate a unified description of radiation-matter-dark energy universe
within the  brane cosmology.
It is confirmed that there is no phantom phase from brane-bulk interactions (HDE-5DCDM, CDM-5DCDM) and
interaction on the brane (HDE-CDM) when using $\omega_{\rm \Lambda}^{\rm eff}$.
Thus our results favors Setare's case~\cite{Setare2} but disfavors Cai-Gong-Wang's case~\cite{CGW}.
This arises mainly because we used a different definition for the effective EoS
 $\omega_{\rm \Lambda}^{\rm eff}$ from Cai-Gong-Wang's case of $\omega_{\rm de}^{\rm eff}$
as well as the HDE as  dark energy.
Recently, the authors in~\cite{WGWA} showed that the interacting holographic dark energy with CDM may lead to
the phantom phase using the native EoS $\omega_{\rm \Lambda}$. Also the authors in~\cite{BNPT}
showed that the brane-bulk interaction without the holographic dark energy accommodates the $\omega=-1$ crossing
when using $\omega_{\rm de}^{\rm eff}$. Hence, the issue is to choose an appropriate EoS for
describing the dark energy universe.

Also, we obtain an additional  information from the unified picture of interactions.
We suggest a sequence of the evolution: radiation-dominated universe $\to$ matter-dominated universe
$\to$ dark energy-dominated universe. The 5DCDM plays the same role as a  radiation on the brane.
As is shown Fig.1 and Table 1 , we have $z_{\rm eq}=27.1$
which is not close to $z^{\rm ob}_{\rm eq}=2.4 \times \Omega_{\rm m}h^2 \simeq 4.8 \times 10^{3}$
 if there is no interaction.
Interestingly, as is shown Fig. 4e and Table 1, there is a good value of $z_{\rm eq}=1.1 \times 10^{3}$,
which is the same order
as  the observational value $z^{\rm ob}_{\rm eq}$
if the interaction between CDM and 5DCDM is included.
However, we do not resolve the coincidence problem because there is no interaction between HDE and CDM.

We stress that if one uses 5D CDM $\chi$ in the brane cosmology instead of radiation,
 its late time evolution  is not sizably
different from the  FRW universe with radiation-matter-dark energy.

Concerning the type of decay rate $\Gamma$, we find that (3)-type is suitable for all  interactions
and thus it could be  regarded as the general form. (2)-type works for three cases of HDE$\to$CDM,
CDM$\to$HDE, and 5DCDM$\to$HDE. Finally, (1)-type works for HDE$\to$CDM only and it belongs to a very restricted
decay rate.

\section*{Acknowledgment}
 K. Kim and H. Lee were  in part supported by
KOSEF, Astrophysical Research Center for the Structure and
Evolution of the Cosmos at Sejong University. Y. Myung  was in
part supported by the Korea Research Foundation
(KRF-2006-311-C00249) funded by the Korea Government (MOEHRD).

\end{document}